\documentclass[12pt]{article}
\usepackage{latexsym}
\usepackage{amsmath,,calrsfs}
\usepackage{amsfonts}
\usepackage{amssymb}
\usepackage{amscd}
\usepackage{bbm}
\usepackage{fancybox}
\usepackage{cite}
\usepackage{amsmath,amsfonts,amsbsy}
\usepackage{pstricks,pst-node}
\usepackage[small,bf,hang]{caption2}
\usepackage{graphicx}
\usepackage{epsfig}
\usepackage{psfrag}
\usepackage{comment}
\usepackage{hyperref}

\usepackage{float}

\psset{unit=1.3cm,linewidth=.5pt,radius=.2}  

\usepackage{multirow}                     
\usepackage{float}                          
\usepackage{lscape}                         
\usepackage{bm}

\newcommand{\beq}{\begin{equation}}
\newcommand{\eeq}{\end{equation}}
\newcommand{\bi}{\begin{itemize}}
\newcommand{\ei}{\end{itemize}}
\newcommand{\bt}{\begin{tabular}}
\newcommand{\et}{\end{tabular}}
\newcommand{\bc}{\begin{center}}
\newcommand{\ec}{\end{center}}

\newcommand{\be}{\begin{equation}}
\newcommand{\ee}{\end{equation}}
\newcommand{\bea}{\begin{eqnarray}}
\newcommand{\eea}{\end{eqnarray}}
\newcommand{\ba}{\begin{array}}
\newcommand{\ea}{\end{array}}

\def\bbox{{\,\lower0.9pt\vbox{\hrule \hbox{\vrule height 0.2 cm
\hskip 0.2 cm \vrule height 0.2 cm}\hrule}\,}}
\newcommand{\dsl}{\pa \kern-0.5em /}

\makeatletter \@addtoreset{equation}{section} \makeatother

\def\slashchar#1{\setbox0=\hbox{$#1$}           
   \dimen0=\wd0                                 
   \setbox1=\hbox{/} \dimen1=\wd1               
   \ifdim\dimen0>\dimen1                        
      \rlap{\hbox to \dimen0{\hfil/\hfil}}      
      #1                                        
   \else                                        
      \rlap{\hbox to \dimen1{\hfil$#1$\hfil}}   
      /                                         
   \fi}

\begin{document}

\begin{titlepage}
\begin{center}

\hfill  DAMTP-2013-46

\vskip 1.5cm

{\Large \bf An $O(D,D)$ Invariant Hamiltonian Action for the Superstring}

\vskip 1cm

{\bf Chris D. A. Blair, Emanuel Malek and Alasdair J. Routh} \\

\vskip 25pt

{\em  Department of Applied Mathematics and Theoretical Physics,\\ Centre for
Mathematical Sciences, University of Cambridge,\\
Wilberforce Road, Cambridge, CB3 0WA, U.K.\vskip 5pt }

{email: {\tt C.D.A.Blair@damtp.cam.ac.uk, E.Malek@damtp.cam.ac.uk, A.J.Routh@damtp.cam.ac.uk}} \\

\end{center}

\vskip 0.5cm

\begin{center} {\bf ABSTRACT}\\[3ex]
\end{center}

\noindent 
We construct $O(D,D)$ invariant actions for the bosonic string and RNS superstring, using
Hamiltonian methods and ideas from double field theory. In this framework 
the doubled coordinates of double field theory appear as coordinates
on phase space and T-duality becomes a canonical transformation. Requiring the
algebra of constraints to close leads to the section condition, which splits the
phase space coordinates into spacetime coordinates and momenta.

\end{titlepage}

\newpage
\setcounter{page}{1}

\newpage

\section{Introduction}

The usual string worldsheet actions have several manifest symmetries, such as invariance under
worldsheet reparametrisations and invariance under the spacetime Poincar\'e group or a curved space
analog. In contrast, T-duality is a hidden symmetry of the action. To better understand this
symmetry, one is led to search for alternative string worldsheet actions, which are written in terms of objects covariant under the $O(D,D)$ transformations of T-duality.

The development of actions with this goal in mind has a long history, with the foundations being laid in 
\cite{Duff:1989tf,Tseytlin:1990nb, Tseytlin:1990va,Siegel:1993th,Siegel:1993xq} and a recent renewal
of interest including \cite{Hull:2004in, Dabholkar:2005ve,Hull:2006va,Hull:2006qs,Berman:2007vi,
Hull:2007jy,DallAgata:2008qz, Hull:2009sg, Nibbelink:2012jb,Lee:2013hma, Nibbelink:2013zda}. Two recent reviews dealing with this material and its extensions are \cite{Aldazabal:2013sca, Berman:2013eva}.
A common feature of these attempts is the introduction of new ``dual coordinates'' in order to place
momentum and winding on an equal footing, as required by T-duality. Momentum is dual to the
spacetime coordinates, $X^{\mu}$, so winding should be dual to some new dual coordinates, $Y_{\mu}$.
With these extra coordinates the string is thought of as moving in a space with twice the usual
number of dimensions, and the result is the \emph{doubled formalism} of the string.

In this doubled formalism the usual coordinates and their duals are packaged into an $O(D,D)$
vector. If the string moves in a background involving a metric, $G$, and Kalb-Ramond field, $B$, one
can place these two fields into a ``generalised metric'', an $O(D,D)$ tensor. The $O(D,D)$
transformations of this object match the well-known Buscher
transformations\cite{Buscher:1987sk,Buscher:1987qj}. Duality invariant actions can then be
constructed from these two $O(D,D)$ tensors. These actions must be supplemented with a constraint
added by hand in order to allow us to (locally) reduce back down to $D$ physical coordinates.

The doubled formalism successfully provides a string worldsheet action with manifest T-duality invariance. The cost of doing this is of course the introduction of extra coordinates, whose precise physical significance is perhaps unclear. In addition, one would like to understand the origin of the actions of doubled formalism actions.

The geometry that is appropriate to understanding the doubled formalism is ``generalised geometry'',
introduced by Hitchin and Gualtieri \cite{Hitchin:2004ut, Gualtieri:2003dx}. In generalised geometry
one extends the tangent bundle to the direct sum of the tangent and cotangent bundles. This can be
interpreted as the tangent bundle in the doubled formalism. 

Generalised geometry is fundamental for the study of T-duality covariance from the spacetime
perspective \cite{Siegel:1993th,Siegel:1993xq,Hull:2009mi, Hull:2009zb, Hohm:2010jy, Hohm:2010pp,
Coimbra:2011nw}. In the ``double field theory'' approach this is achieved by
doubling the number of coordinates. This leads to a rewriting of the low energy effective action of the string in
terms of the above $O(D,D)$ vector of coordinates and their duals, and $O(D,D)$ tensors, giving a
manifest $O(D,D)$ covariant form of the string's low energy theory. In the bosonic case, at least, this action is the low-energy effective action of the \emph{doubled} string \cite{Berman:2007xn,Copland:2011yh,Copland:2011wx}. Double field theory has also been applied with success to the 10-dimensional supergravities
\cite{
Jeon:2010rw, Jeon:2011kp, Jeon:2011cn, Jeon:2011vx,
Jeon:2011sq,Jeon:2012kd,Jeon:2012hp,Hohm:2010xe, Hohm:2011ex, Hohm:2011zr, Hohm:2011dv,  Hohm:2011nu
}
and allows important insights into string theory beyond its supergravity regime, such as non-geometric backgrounds, as for example discussed in the reviews \cite{Aldazabal:2013sca, Berman:2013eva}.

In double field theory the fields can at first depend on an $O(D,D)$ vector of coordinates. However,
just as in the doubled formalism where half the coordinates must be eliminated by a constraint, to
make contact with the usual low energy theories half the dependence of the fields in double field
theory must be eliminated. This is achieved by the ``section condition''. The coordinates of double
field theory can be thought of as parametrising a doubled space with some unusual properties
\cite{Hohm:2011si,Hohm:2012gk, Hohm:2012mf,Park:2013mpa,Hohm:2013bwa}. The local symmetries of this space form ``generalised diffeomorphisms'' \cite{Hull:2009zb,Berman:2012vc}, and the section condition arises by requiring that their algebra closes. In this paper, we will show how the section condition also appears when studying the symmetry algebra of the \emph{worldsheet} diffeomorphisms of a doubled string.

Another profitable approach to T-duality is to treat the string from the Hamiltonian viewpoint, since
the Hamiltonian of a string can naturally be expressed in an $O(D,D)$ invariant form. The spacetime
momentum, $P_\mu$, and the worldsheet spatial derivatives of the spacetime coordinates, $X'^{\mu}$,
combine naturally into an $O(D,D)$ vector. The Hamiltonian can then be written in terms of this
vector and the generalised metric in a manifestly $O(D,D)$ covariant way. In this framework,
T-dualities appear as canonical transformations. This point has been stressed recently by Maharana
\cite{Maharana:2010sp, Maharana:2013uvy, Maharana:2012fv}, while a phase space T-duality manifest
action has also appeared in \cite{Freidel:2013zga}.

This paper has two aims. Firstly, to combine insights from both the Hamiltonian and doubled approaches to T-duality to present a new way to construct $O(D,D)$ invariant string actions. By making use of the Hamiltonian framework we hope to clarify the relationship between the doubled approach and the normal formulation of the string, particularly the status of the extra coordinates in the former. Secondly, to construct a doubled action for the RNS superstring in a non-trivial $G$ and $B$ background.

Different but related approaches to the study of the doubled string from a Hamiltonian perspective include \cite{HackettJones:2006bp, West:2010rv}. 

We begin in section 2 with the $\sigma$-model for the bosonic string moving through a curved background with a Kalb-Ramond field. Since the Hamiltonian of the string is naturally $O(D,D)$ invariant, we rewrite the action in ``Hamiltonian form'',
\be
S = \int d\tau d\sigma \left( \dot{X}\cdot P - {\rm Ham}(X,P)\right) \,,
\ee
where ${\rm Ham}$ denotes the Hamiltonian. In order to make contact with double field theory and assemble everything into $O(D,D)$ tensors we introduce the dual coordinates $Y_\mu$. This is done by making the replacement $P_\mu \mapsto Y'_{\mu}$, which has been suggested by previous work and can be justified by T-duality. The $2D$ doubled coordinates of double field theory are therefore interpreted as the $2D$ coordinates on the phase space of a $D$ dimensional system. This action naturally defines a constrained Hamiltonian system, which we analyse. We find the section condition of double field theory as a requirement for the closure of the constraint algebra, or equivalently for worldsheet reparametrisation invariance.

In section 3 we analyse the RNS superstring in the same way. Starting with the corresponding $\sigma$-model, we construct the Hamiltonian form of the action and then an $O(D,D)$ invariant form as for the bosonic string. Unlike the spacetime coordinates, $X^{\mu}$, the fermions do not get ``doubled''; this is consistent with the phase space interpretation of the bosonic doubled coordinates. We analyse the algebra of constraints and again find the section condition.

\section{The Bosonic String}

The $\sigma$-model action for a bosonic string propagating in a $D$-dimensional
background with metric $G_{\mu \nu}(X)$ and Kalb-Ramond field $B_{\mu \nu}(X)$
is\footnote{$\mu, \nu, \ldots$ are spacetime indices, $\alpha, \beta, \ldots$ are worldsheet indices, the worldsheet metric has signature $(-,+)$ and $\epsilon^{01} = -1$ and we have set the string tension to $1$.}
\be \label{eq:bosonicsigmamodel}
S = -\frac{1}{2} \int d\tau d\sigma  \left( \sqrt{-h}h^{\alpha
\beta}\partial_{\alpha}X^{\mu} \partial_{\beta}X^{\nu} G_{\mu \nu}(X) +
\epsilon^{\alpha \beta}\partial_{\alpha}X^{\mu} \partial_{\beta}X^{\nu} B_{\mu
\nu}(X) \right) \,.
\ee
We will begin by putting this action into Hamiltonian form. It will be convenient to use a general parametrisation of the worldsheet
metric in terms of three real parameters, $\lambda$, $\tilde\lambda$ and $\Omega$:
\be
h_{\alpha \beta} = \Omega \left(\begin{array}{cc} -\lambda\tilde{\lambda} & \frac{1}{2}(\tilde{\lambda}-\lambda) \\ \frac{1}{2}(\tilde{\lambda} - \lambda) & 1\end{array}\right)\,,
\ee
a parametrisation which will be justified when we find the Hamiltonian. The conformal scale $\Omega$ drops out of the action as expected. With this parametrisation, the momentum conjugate to $X^{\mu}$ is
\be
P_{\mu} \equiv \frac{\partial L}{\partial \dot{X}^{\mu}} =
\frac{2}{\lambda + \tilde{\lambda}}G_{\mu\nu}\left(\dot{X}^{\nu} + \frac{\lambda - \tilde{\lambda}}{2} X^{\prime \nu}\right) + B_{\mu \nu}X^{\prime
\nu} \,.
\ee
We can Legendre transform to find the Hamiltonian
\be
{\rm Ham}(X,P) = \frac{\lambda}{4}(P - (G+B)X')^2 + \frac{\tilde{\lambda}}{4}(P + (G-B)X')^2 \,,
\ee
suppressing indices and using $G_{\mu\nu}$ to square. Like in any reparametrisation invariant theory, the Hamiltonian is a sum of constraints with corresponding Lagrange multipliers.

To make this Hamiltonian manifestly duality invariant we need to package all the fields into representations of $O(D,D)$. We can construct an $O(D,D)$ vector out of $X'^{\mu}$ and $P_{\mu}$
\be
Z^M = \left(\begin{array}{c}X'^{\mu} \\ P_{\mu}\end{array}\right)
\ee
and we package the background fields into the generalised metric, $H_{MN}$. We will also need to use the defining $O(D,D)$ form $\eta_{MN}$. These two tensors can be written
\be
H_{MN} = \left(\begin{array}{cc}G - BG^{-1}B & BG^{-1} \\ -G^{-1}B &
G^{-1}\end{array}\right) \,, \qquad \eta_{MN} = \left(\begin{array}{cc}0 &
\mathbb{I} \\ \mathbb{I} & 0\end{array}\right) \,.
\ee
These will naturally combine into the projectors
\begin{equation}
 \Pi_{MN} = \frac{1}{2} \left( H_{MN} - \eta_{MN} \right) \,, \qquad \tilde{\Pi}_{MN} = \frac{1}{2} \left( H_{MN} + \eta_{MN} \right). \footnotemark
\end{equation}
\footnotetext{Throughout this paper we will use the convention that $O(D,D)$ indices are lowered and
raised by the generalised metric, $H_{MN}$, not $\eta_{MN}$. This means in particular that
$\Pi_M{}^N = - \Pi_{ML} \eta^{LN}$, whereas $\tilde\Pi_{M}{}^N = \tilde\Pi_{ML} \eta^{LN}$.}
\noindent Under the action of $O(D,D)$, the transformation of the generalised metric exactly matches the Buscher transformations of the background fields. We can now write the Hamiltonian entirely in terms of $O(D,D)$ covariant objects
\be
{\rm Ham}(X,P) = \frac{\lambda}{2}Z^M \Pi_{MN}Z^N + \frac{\tilde{\lambda}}{2}Z^M \tilde{\Pi}_{MN}Z^N \,.
\ee
We will use this $O(D,D)$ invariant Hamiltonian to construct an $O(D,D)$ invariant action. By the definition of the Hamiltonian, the Lagrangian is equal to
\be
\mathcal{L} = \dot{X}\cdot P - {\rm Ham}(X,P)\,,
\ee
and we can integrate over the worldsheet to get the action in Hamiltonian form:
\be \label{eq:hamiltonianaction1}
S = \int d\tau d\sigma \, \left( \dot{X}\cdot P - \frac{\lambda}{4}(P - (G+B)X')^2 -
\frac{\tilde{\lambda}}{4}(P + (G-B)X')^2\right)\,.
\ee
The equivalence of this action to \eqref{eq:bosonicsigmamodel} can be checked by eliminating $P_{\mu}$ via its equation of motion.

In this formulation it is most natural to study the dynamics as of the string as a constrained Hamiltonian system. The dynamical fields are the
canonical pair of $D$-vectors $(X^{\mu}, P_{\mu})$ and the constraints are
\be
\mathcal{H} = \frac{1}{4}\left(P - (G+B)X'\right)^2 \,, \qquad \tilde{\mathcal{H}} =
\frac{1}{4}\left(P + (G-B)X'\right)^2 \,,
\ee
which generate reparametrisations in the lightcone directions $\tau - \sigma$ and $\tau + \sigma$ respectively. These constraints are imposed by
Lagrange multipliers $\lambda$ and $\tilde{\lambda}$, the sum and difference of
which are, up to factors of the string tension, the ``lapse'' and ``shift'' familiar from the
Hamiltonian formulation of general relativity.

All $\tau$ derivatives appear in the first term of the action, and from this we can read off the fundamental Poisson bracket
\be
\{ X^{\mu}(\sigma_1), P_{\nu}(\sigma_2) \} = \delta^{\mu}_{\nu} \delta(\sigma_1
- \sigma_2) \,,
\ee
while all others vanish. Calculations are made easier by working with ``smeared'' constraints, defined for any constraint $C(\sigma)$ and test function $\alpha$ by
\be
C(\alpha) \equiv \int d\sigma \, \alpha(\sigma) C(\sigma) \,.
\ee
In this notation the algebra of the smeared constraints can be worked out to be
\be
\begin{split}
\{ \mathcal{H}(\alpha), \mathcal{H}(\beta) \} &= -
\mathcal{H}(\alpha\beta'-\beta\alpha') \,, \\
\{ \mathcal{H}(\alpha), \tilde{\mathcal{H}}(\beta) \} &= 0 \,, \\
\{ \tilde{\mathcal{H}}(\alpha), \tilde{\mathcal{H}}(\beta) \} &=
\tilde{\mathcal{H}}(\alpha\beta'-\beta\alpha') \,.
\end{split}
\ee
We recognise this as the $\rm{Diff}_1 \times \rm{Diff}_1$ algebra which ensures that the action is worldsheet diffeomorphism invariant. To make contact with more familiar formulations of the string, note that the two constraints can be Fourier expanded in $\sigma$, and the modes of each constraint would give an independent Virasoro algebra.

In the usual doubled formalism of the string \cite{Hull:2004in}, $X^{\mu}$ is packaged into an $O(D,D)$ vector with a set of dual coordinates, $Y_{\mu}$
\be
X^M = \left(\begin{array}{c}X^{\mu} \\ Y_{\mu}\end{array}\right) \,.
\ee
The Hamiltonian tells us that $X'^{\mu}$ is naturally packaged with $P_{\mu}$, so we should investigate the formal replacement\footnote{It is actually $TX'$ which appears in an $O(D,D)$ vector with $P$, so when the tension is restored, $P$ should be set to $TY'$.}
\be
P_{\mu} \mapsto Y'_{\mu} \,.
\ee
which changes the Poisson bracket to
\be
\label{eq:XYbracket}
\{ X^{\mu}(\sigma_1), Y'_{\nu}(\sigma_2) \} = \delta^{\mu}_{\nu} \delta(\sigma_1
- \sigma_2) \,.
\ee
The $O(D,D)$ covariant extension of this bracket would be
\be
\label{eq:XXbracket}
\{ X^M(\sigma_1), X'^N(\sigma_2) \} = \eta^{MN} \delta(\sigma_1
- \sigma_2) \,,
\ee
which requires that
\be
\label{eq:XYbracket2}
\{ Y_{\mu}(\sigma_1), X'^{\nu}(\sigma_2) \} = \delta^{\nu}_{\mu} \delta(\sigma_1
- \sigma_2) \,.
\ee
However if we first differentiate \eqref{eq:XYbracket} with respect to $\sigma_2$ to obtain
\be
\{ Y'_{\mu}(\sigma_1), X'^{\nu}(\sigma_2) \} = \delta^{\nu}_{\mu} \delta'(\sigma_1
- \sigma_2) 
\ee
and then integrate over $\sigma_1$ we find that \eqref{eq:XYbracket2} holds up to a constant of integration
\be
\{ Y_{\mu}(\sigma_1), X'^{\nu}(\sigma_2) \} = \delta^{\nu}_{\mu} \left(\delta(\sigma_1
- \sigma_2) - C \right) \,.
\ee
Now we note that the zero-mode of $Y_{\mu}$ is absent from the action and thus has zero bracket with
everything. If we integrate both sides with respect to $\sigma_1$, we find that $C = -\frac{1}{2\pi}$,
exactly cancelling the zero-mode of the $\delta$-function, so we do not obtain the covariant bracket
\eqref{eq:XYbracket2}. We can understand this as follows: the zero-mode
of $X^{\mu}$ appears in the action, but the zero-mode of $Y_{\mu}$ does not, breaking $O(D,D)$ covariance. We will now investigate the consequences of including a term involving the zero-mode of $Y_{\mu}$ in the action.

It is helpful to split the variables up into zero-mode, winding and oscillator parts. To implement winding, we demand the boundary condition $X^{\mu}(2\pi) - X^{\mu}(0) = w^{\mu}$, where $w^{\mu}$ is a constant, non-dynamical vector. We can define the zero-modes
\be
x^{\mu} = \int d\sigma \, X^{\mu} \,, \qquad p_{\mu} = \frac{1}{2\pi} \int d\sigma \, P_{\mu} \,,
\ee
and then write
\be
X^{\mu} = x^{\mu} + \frac{1}{2 \pi} w^\mu ( \sigma -  \pi) + \bar{X}^{\mu} \,, \qquad P_{\mu} = \frac{1}{2\pi}p_{\mu} + \bar{P}_{\mu} \,.
\ee
The barred quantities are periodic and integrate to zero, so can be expressed as Fourier series in $\sigma$. The quantity $\bar{Y}$ such that $\bar{Y}' = \bar{P}$ and $\int d\sigma \, \bar{Y} = 0$ is therefore well-defined. We can define $Y'$ via
\be
Y'_{\mu} = P_{\mu} = \frac{1}{2\pi}p_{\mu} + \bar{Y}'_{\mu} \,.
\ee
This is so far just a renaming of variables, but we can integrate this with respect to $\sigma$, picking up a constant of integration,
\be \label{Ydef}
Y_{\mu} = y_{\mu} + \frac{1}{2\pi}p_{\mu}(\sigma - \pi) + \bar{Y}_{\mu} \,.
\ee
This introduces an extra zero-mode to the theory, but this has no effect as only $Y'$ appears in the action. The Hamiltonian is written in terms of $Z^M = X'^M$, and hence in terms of $O(D,D)$ objects, but the first term requires more work. Consider
\be
\int d\sigma \, \dot{X} \cdot P = \dot{x} \cdot p + \int d\sigma \, \dot{\bar{X}} \cdot \bar{P} \,.
\label{eq:intdotXP}
\ee
We notice that
\be
\int d\tau d\sigma \, \dot{\bar{X}} \cdot \bar{P} = \int d\tau d\sigma \, \left( \frac{1}{2} \dot{\bar{X}} \cdot \bar{Y}' + \frac{1}{2} \bar{X}' \cdot \dot{\bar{Y}} + \frac{1}{2}\frac{d}{d\tau}(\bar{X}\cdot \bar{Y}') - \frac{1}{2}\frac{d}{d\sigma}(\bar{X}\cdot \dot{\bar{Y}}) \right) \,.
\ee
By periodicity in $\sigma$ we can drop the last term, and we are also free to drop the total $\tau$ derivative, so that
\be
\int d\tau d\sigma \, \dot{X} \cdot P = \dot{x} \cdot p + \int d\tau d\sigma \, \frac{1}{2} \dot{\bar{X}}^M \eta_{MN} \bar{X}'^N  \,.
\ee
Everything is now $O(D,D)$ covariant except for the zero-mode term $\dot{x}\cdot p$. To deal with
this, we note from equation \eqref{Ydef} that $p_{\mu}$ can be interpreted as a winding in the
$Y_{\mu}$ direction. Just as the momentum $p_{\mu}$ is conjugate to position $x^{\mu}$, the winding
$w^{\mu}$ should be conjugate to $y_{\mu}$. Furthermore, for $O(D,D)$ covariance $w^{\mu}$ should be put on the same footing as $p_{\mu}$, and so be a dynamical variable. The zero-modes naturally combine into two $O(D,D)$ vectors
\be
x^M = \left(\begin{array}{c}x^{\mu} \\ y_{\mu}\end{array}\right) \,, \qquad p_M =
\left(\begin{array}{c}p_{\mu} \\ w^{\mu}\end{array}\right) \,,
\ee
and we replace the term $\dot{x} \cdot p$ with the fully $O(D,D)$ invariant $\dot{x}^M p_M$.

To reiterate, we have made two modifications to the action. We have promoted $w^{\mu}$ to a
dynamical variable, and we have added a term $\dot{y} \cdot w$ to the action. As $y_{\mu}$ only
appears in this term, the equation of motion of $y$ sets $\dot{w} = 0$, and the equation of motion
of $w$ sets the new variable $y$ to some function of the other variables. The equations of motion of
the new action are identical to the equations of motion of the original action with non-dynamical
$w$ plus an equation determining $\dot{y}$ which does not affect the other variables. The actions are therefore classically equivalent, although there may be further implications quantum mechanically. This issue of zero-modes has been considered by other authors in \cite{Hull:2006va, Berman:2007vi}. 

Our final action is therefore
\be
\label{eq:bosonicoddaction}
S = \int d\tau d\sigma \, \left( \frac{1}{2\pi}\dot{x}^M p_M + \frac{1}{2}\dot{\bar{X}}^M \eta_{MN} \bar{X}'^N - \frac{\lambda}{2}X'^M \Pi_{MN} X'^N - \frac{\tilde{\lambda}}{2}X'^M \tilde{\Pi}_{MN} X'^N \right) \,.
\ee
The form of this action is a little strange, given the separate treatment of zero-modes and
oscillators, but the extra term is exactly that needed for the natural bracket \eqref{eq:XXbracket},
and to treat momentum and winding equally. We stress again that that the normal string action, with boundary conditions involving non-dynamical
winding, is equivalent to the doubled action \eqref{eq:bosonicoddaction} where winding is allowed to
be dynamical. We will later see that this action not only describes configurations with constant
winding but, after T-duality transformations, situations with non-constant winding along isometry
directions.

Up to this point the background fields, expressed either as the generalised metric, $H$, or as $G$
and $B$, have been functions of $X^{\mu}$ only. This is not $O(D,D)$ covariant. However, allowing
the fields to depend on the full $X^M$ would be the same as allowing the fields of the normal string
to depend on both the spacetime coordinates and momenta. If one were to then compute the algebra of
constraints, extra terms proportional to the momentum derivatives of the background fields would
appear, preventing the algebra from closing. The momentum independence of the background fields is
therefore required for consistency.

We will temporarily allow the generalised metric to depend on the full $X^M$, and find that similar extra terms appear in the algebra, preventing its
closure. Requiring that these terms vanish will give us an analogous consistency condition. Crucially, this condition will be $O(D,D)$ invariant.

With this extra step, the usual string action
\eqref{eq:hamiltonianaction1} and the manifestly $O(D,D)$ invariant action
\eqref{eq:bosonicoddaction}
 differ only by the classically irrelevant zero-mode term and the covariant way in which momentum dependence of the background fields is
prevented.

We now analyse \eqref{eq:bosonicoddaction} as a constrained Hamiltonian system. Denoting by $\bar{P}_M$ the momentum conjugate to $\bar{X}^M$, we find a set of second-class constraints,
\be
C_M = \bar{P}_M - \frac{1}{2} \eta_{M N} \bar{X}^{\prime N} \,.
\ee
We pass to (a priori) partial Dirac brackets in these second-class constraints, finding as promised the integrated version of \eqref{eq:XXbracket}
\be
\{X^M(\sigma_1), X^N(\sigma_2)\}^* = -\eta^{MN} \epsilon(\sigma_1 -
\sigma_2) \,,
\label{eq:DiracBrackets}
\ee
where $\epsilon$ is such that $\epsilon' = \delta$, the Dirac delta. The remaining constraints are
\be
\mathcal{H} = \frac{1}{2} X'^M\Pi_{MN}X'^N \,, \qquad \tilde{\mathcal{H}} = \frac{1}{2}
X'^M \tilde\Pi_{MN}X'^N \,.
\ee
We can now use the brackets \eqref{eq:DiracBrackets} to compute the algebra of the remaining constraints,
finding
\be
\begin{split}
\{ \mathcal{H}(\alpha), \mathcal{H}(\beta) \}^* &= -
\mathcal{H}(\alpha\beta'-\beta\alpha') + \Delta \,, \\
\{ \mathcal{H}(\alpha), \tilde{\mathcal{H}}(\beta) \}^* &= \Delta \,, \\
\{ \tilde{\mathcal{H}}(\alpha), \tilde{\mathcal{H}}(\beta) \}^* &=
\tilde{\mathcal{H}}(\alpha\beta'-\beta\alpha') + \Delta \,.
\end{split}
\ee
This is the $\rm{Diff}_1 \times \rm{Diff}_1$ algebra up to the extra term
\be
\Delta = \frac{1}{16} \int d\sigma_1 d\sigma_2 \, \alpha(\sigma_1)
\beta(\sigma_2) \, X'^M(\sigma_1) X'^N(\sigma_1) X'^P(\sigma_2) X'^Q(\sigma_2)
\{H_{MN}(\sigma_1), H_{PQ}(\sigma_2)\}^* \,.
\ee
As discussed earlier, the appearance of this term is a direct consequence of allowing the
generalised metric to depend on the full $X^M$. It therefore tells us how to restrict the dependence
in an $O(D,D)$ covariant way. Note that
\be
\{H_{MN}(\sigma_1), H_{PQ}(\sigma_2)\}^* = -\eta^{RS}\partial_R H_{MN}(\sigma_1) \partial_S H_{PQ}(\sigma_2) \epsilon(\sigma_1 - \sigma_2) \,,
\ee
and so a sufficient condition for $\Delta$ to vanish, and for all brackets with $\Delta$ in to vanish is
\be
\eta^{RS}\partial_R F_1(\sigma_1) \partial_S F_2(\sigma_2) = 0 \qquad \forall \sigma_1, \sigma_2 \,,
\ee
where $F_1$ and $F_2$ are any combinations of $H_{MN}$ and its spacetime derivatives. We recognise
this as the section condition of double field theory in a form studied previously in \cite{Copland:2011wx}; it
essentially requires the generalised metric to only depend on at most half of the coordinates. 
The consistency of this condition will have to be checked upon quantisation. It is possible that a slightly weaker condition on the generalised metric could cause
$\Delta$ to vanish
\cite{Aldazabal:2011nj,Geissbuhler:2011mx,Grana:2012rr,Berman:2013cli,Geissbuhler:2013uka, Berman:2013uda}, which should be studied in the future; in this paper we will impose the section
condition in its full strength. This condition should be applied before performing the Hamiltonian
analysis, although in this case it does not make any difference to the result.

Note that we are not treating $\Delta$ as a new constraint in the Hamiltonian sense. We are viewing it as a condition on the admissible
background fields, not on the dynamical variables. This is analogous to requiring, for the normal
string action,
\be
\frac{\partial G_{\mu\nu}}{\partial P_{\rho}} = 0 \,,
\ee
which one would certainly not treat as a constraint.

There are other reasons why we do not treat $\Delta$ as a constraint on the dynamical
variables. 
Firstly, the algebra of constraints would no longer be the
$\rm{Diff}_1 \times \rm{Diff}_1$ algebra, breaking the reparametrisation invariance of the string. 
Secondly, adding additional constraints would change the number of physical degrees of
freedom, which we do not want to do. 

After imposing the section condition, the remaining constraints, $\mathcal{H}$ and
$\mathcal{\tilde{H}}$, are first-class with respect to the Dirac brackets \eqref{eq:DiracBrackets}.
These brackets are thus actually full Dirac brackets and we have completed our analysis of
constraints.

The section condition allows us to reduce our action \eqref{eq:bosonicoddaction} back to the
``undoubled'' action \eqref{eq:hamiltonianaction1}. 
It implies that we can go to a frame
\be
X^M = \left(\begin{array}{c}X^{\mu} \\ Y_{\mu}\end{array}\right) \,,
\ee
where the generalised metric is independent of $Y_{\mu}$. The second term in
\eqref{eq:bosonicoddaction} can be integrated by parts so that it becomes
$\dot{\bar{X}}^{\mu}\bar{Y}'_{\mu}$, and the $\dot{y}\cdot w$ can be ignored classically if $w^{\mu}$ is set constant. As $Y_{\mu}$ then appears in the action only as $Y'_{\mu}$ we may rename $Y'_{\mu} \mapsto P_{\mu}$ and recover the undoubled action.

In fact our action can also describe non-constant winding by picking a different frame. We may pick
as one of the $X^\mu$ a direction upon which the generalised metric does not depend, in other words
an isometry direction. The winding in this direction need not be constant, but the momentum, the
winding of the dual direction, must be. 

We can also relate our action \eqref{eq:bosonicoddaction} to a form of doubled action used in the
literature. By going to conformal gauge, $\lambda = \tilde{\lambda} = 1$, we
obtain essentially the original doubled action introduced by Tseytlin in \cite{Tseytlin:1990nb}. The
only difference lies in our treatment of the zero modes which is essential to obtain the expected Dirac
bracket. 

The first-class constraints generate gauge transformations associated to the diffeomorphism symmetry which the action enjoys. Under a gauge transformation $X^M$ transforms as
\be
\begin{split}
\left\{ \mathcal{H}(\alpha) , X^M(\sigma) \right\}^* &= \alpha \Pi^M{}_N X'^N + \frac{1}{4}\int d\sigma' \, (\alpha X'^PX'^Q\eta^{MN}\partial_N H_{PQ})(\sigma')\epsilon(\sigma' - \sigma) \, , \\
\left\{ \tilde{\mathcal{H}}(\alpha) , X^M(\sigma) \right\}^* &= -\alpha \tilde{\Pi}^M{}_N X'^N + \frac{1}{4}\int d\sigma' \, (\alpha X'^PX'^Q\eta^{MN}\partial_N H_{PQ})(\sigma')\epsilon(\sigma' - \sigma) \, .
\end{split}
\ee
The gauge transformations induce non-local contributions, but this problem is also solved by the
section condition. If the generalised metric, $H_{MN}$, depends on a coordinate $X_*$, then its $\eta_{MN}$-dual coordinate $Y^*$ will transform non-locally. If we want the generalised metric to depend only on coordinates which transform locally on the worldsheet, then it can depend on at most $D$ coordinates and must be independent of their duals. These coordinates then transform locally, and their duals only appear in the action through their $\sigma$-derivatives, which are also local.

\section{The RNS Superstring}

The $\sigma$-model action for an RNS superstring propagating in a $D$-dimensional
background with metric $G_{\mu \nu}(X)$ and Kalb-Ramond field $B_{\mu \nu}(X)$ is \cite{Bergshoeff:1985qr} \footnote{The flat $\gamma$-matrices are $\gamma^0 = i\sigma_2$, $\gamma^1 = -\sigma_1$ and $\gamma_3 = -\sigma_3$, and $\bar{\chi} = \chi^T \gamma^0$. $T_{\mu \nu \rho} = 3 \partial_{[\mu} B_{\nu \rho]}$ is the field strength of the Kalb-Ramond field, $\Gamma^\mu{}_{\nu \rho}$ is the Levi-Civita connection of the metric and the associated Riemann tensor in our conventions is $R^\mu{}_{\nu \rho \sigma} = 2 \partial_{[\rho} \Gamma^\mu{}_{\sigma]\nu} +2 \Gamma^\mu{}_{[\rho| \lambda} \Gamma^\lambda{}_{\sigma] \nu}$.}
\be
\begin{split} \label{eq:rnssigmamodel}
S = -\frac{1}{2} \int &d\tau d\sigma \, \sqrt{-h}\left( h^{\alpha
\beta}\partial_{\alpha}X^{\mu} \partial_{\beta}X^{\nu} G_{\mu \nu} +
\frac{\epsilon^{\alpha \beta}}{\sqrt{-h}}\partial_{\alpha}X^{\mu} \partial_{\beta}X^{\nu} B_{\mu
\nu} \right.  \\
& -i\gamma^{\alpha}\partial_{\alpha}\Psi^{\nu}G_{\mu \nu} - i\bar{\Psi}^{\mu}\gamma^{\alpha}\Psi^{\rho}\Gamma^{\nu}\,_{\rho\sigma}\partial_{\alpha}X^{\sigma}G_{\mu \nu}  -\frac{1}{2}\bar{\Psi}^{\mu}\gamma^{\alpha}\gamma_3\Psi^{\nu}\partial_{\alpha}X^{\rho}T_{\mu\nu\rho}  \\
&+\frac{1}{6}R_{\mu\rho\nu\sigma}\bar{\Psi}^{\mu}\Psi^{\nu}\bar{\Psi}^{\rho}\Psi^{\sigma} + \frac{1}{8}\nabla_{\rho}T_{\mu\sigma\nu}\bar{\Psi}^{\mu}\Psi^{\rho}\bar{\Psi}^{\nu}\gamma_3\Psi^{\sigma}  \\
& -\frac{1}{16}T_{\mu\rho\kappa}T^{\kappa}\,_{\nu\sigma}\bar{\Psi}^{\mu}\gamma_3\Psi^{\rho}\bar{\Psi}^{\nu}\gamma_3\Psi^{\sigma} -2i\bar{\chi}_{\alpha}\gamma^{\beta}\gamma^{\alpha}\Psi^{\mu}\partial_{\beta}X^{\nu}G_{\mu\nu}  \\
& \left. - \frac{1}{6}\bar{\chi}_{\alpha}\gamma^{\beta}\gamma^{\alpha}\Psi^{\mu}\bar{\Psi}^{\nu}\gamma_{\beta}\gamma_3\Psi^{\rho}T_{\mu\nu\rho} + \frac{1}{2}\bar{\chi}_{\alpha}\gamma^{\beta}\gamma^{\alpha}\chi_{\beta}\bar{\Psi}^{\mu}\Psi^{\nu}G_{\mu\nu}\right)\,.
\end{split}
\ee
This extends the bosonic action \eqref{eq:bosonicsigmamodel} by the addition of $\Psi^{\mu}$, a worldsheet two-component Majorana spinor and a spacetime vector, and a worldsheet gravitino, $\chi_{\alpha}$, a worldsheet vector-spinor and spacetime scalar. To derive the Hamiltonian form of this action we use the same parametrisation of the worldsheet metric as before. We also choose a convenient parametrisation of the worldsheet vielbein, so that
\be
h_{\alpha \beta} = \Omega \left(\begin{array}{cc} -\lambda\tilde{\lambda} & \frac{1}{2}(\tilde{\lambda}-\lambda) \\ \frac{1}{2}(\tilde{\lambda} - \lambda) & 1\end{array}\right)\,, \qquad L^{-1} = \Omega^{-\frac{1}{2}}\left(\begin{array}{cc} \frac{2}{\lambda + \tilde{\lambda}} & 0 \\ \frac{\lambda - \tilde{\lambda}}{\lambda + \tilde{\lambda}} & 1\end{array}\right) \,.
\ee
If we rescale the fermions by
\be
\Psi \mapsto \Omega^{-\frac{1}{4}}\Psi \,, \qquad \chi \mapsto \Omega^{\frac{1}{4}}\chi \,,
\ee
then the conformal factor again drops out. It is convenient to split the worldsheet spinor $\Psi^\mu$ into its components as
\be
\Psi^{\mu} = \left(\begin{array}{c}\psi^{\mu}\\ \tilde{\psi}^{\mu}\end{array}\right)\,.
\ee
We also introduce a vielbein $e_\mu{}^a$ in spacetime, so that $G_{\mu \nu} = e_\mu{}^a e_\nu{}^b \eta_{ab}$ where $\eta_{ab}$ is the flat Minkowski metric. We use this vielbein to flatten the indices on the fermions so that $\psi^a = e_{\mu}{}^a \psi^\mu$, $\tilde \psi^a = e_\mu{}^a \tilde\psi^\mu$.

We now follow the same procedure as in the bosonic case. First, we find the momentum conjugate to $X^{\mu}$. We then Legendre transform to find the Hamiltonian and write the action in Hamiltonian form. We do not Legendre transform in the fermionic variables. Again two components of the metric remain and enforce two bosonic constraints; they are now joined by two surviving components of the worldsheet gravitino, which we call $\xi$ and $\tilde{\xi}$, that enforce two new fermionic constraints. The resulting Hamiltonian form action is
\be
S = \int d\tau d\sigma \, \left( \dot{X}\cdot P - \frac{i}{2}(\psi^a\dot{\psi}^b +
\tilde{\psi}^a\dot{\tilde{\psi}}^b)\eta_{ab} -\lambda\mathcal{H} - \tilde{\lambda}\tilde{\mathcal{H}} - i\xi\mathcal{Q} - i\tilde{\xi}\tilde{\mathcal{Q}} \right) \,, \footnotemark
\label{eq:susyhamform}
\ee \footnotetext{Superconformal gauge corresponds to setting $\lambda = \tilde{\lambda} = 1$, $\xi = \tilde{\xi} = 0$.}
where
\be
\begin{split}
4\mathcal{H} &= \left(P_{\mu} - (G+B)_{\mu\nu}X'^{\nu} + \frac{i}{2}\omega_{-\mu a b}\psi^a\psi^b + \frac{i}{2}\omega_{+\mu a b}\tilde{\psi}^a\tilde{\psi}^b\right)^2  \\
& + 2i\left(\eta_{ab}\psi^a\psi'^b + \omega_{-\mu a b}\psi^a\psi^bX'^{\mu}\right) + 2R_{\pm acbd}\tilde{\psi}^a\psi^b\tilde{\psi}^c\psi^d \,,  \\
4\tilde{\mathcal{H}} &= \left(P_{\mu} + (G-B)_{\mu\nu}X'^{\nu} + \frac{i}{2}\omega_{-\mu a b}\psi^a\psi^b + \frac{i}{2}\omega_{+\mu a b}\tilde{\psi}^a\tilde{\psi}^b\right)^2  \\
& - 2i\left(\eta_{ab}\tilde{\psi}^a\tilde{\psi}'^b + \omega_{+\mu a b}\tilde{\psi}^a\tilde{\psi}^bX'^{\mu}\right) + 2R_{\pm acbd}\tilde{\psi}^a\psi^b\tilde{\psi}^c\psi^d \,, \\
2\mathcal{Q} &= \left(P_{\mu} - (G+B)_{\mu\nu}X'^{\nu} + \frac{i}{2}\omega_{-\mu a b}\psi^a\psi^b + \frac{i}{2}\omega_{+\mu a b} \tilde{\psi}^a\tilde{\psi}^b + \frac{i}{6}T_{\mu a b}\psi^a\psi^b\right)\psi^{\mu} \,, \\
2\tilde{\mathcal{Q}} &= \left(P_{\mu} + (G-B)_{\mu\nu}X'^{\nu} + \frac{i}{2}\omega_{-\mu a b}\psi^a\psi^b + \frac{i}{2}\omega_{+\mu a b}\tilde{\psi}^a\tilde{\psi}^b - \frac{i}{6}T_{\mu a b}\tilde{\psi}^a\tilde{\psi}^b\right)\tilde{\psi}^{\mu} \,.
\end{split}
\ee
We have here the following torsionful connections and associated spin connections
\be
\Gamma_{\!\pm}^{\mu}{}_{\nu \rho} =  \Gamma^\mu{}_{\nu \rho} \pm \frac{1}{2} T_\nu{}^\mu{}_\rho\,, 
\quad\quad
\omega_{\pm \mu}{}^a{}_b = e_\nu^a \partial_\mu e^\nu_b + \Gamma^{a}_{\!\pm}{}_{\mu b} \,. \label{eq:tspinconnection2}
\ee
We denote by $R^{\mu}_{\pm}{}_{\nu \rho \sigma}$ the Riemann tensors associated to
$\Gamma^\mu_{\!\pm}{}_{\nu \rho}$. These are in fact equal as a result of the Bianchi identity
$\partial_{[\mu } T_{\nu \rho \sigma]} = 0$. The action \eqref{eq:susyhamform} can of course be converted back into \eqref{eq:rnssigmamodel} by eliminating $P_{\mu}$.

The Poisson brackets for the bosonic variables are as before. The canonical momenta of the fermions are proportional to the fermions themselves; this leads to a set of second-class constraints which must be dealt with by passing to Dirac brackets. This is a standard situation so we just give the results
\be
\begin{split}
\lbrace X^{\mu}(\sigma_1), P_{\nu}(\sigma_2) \rbrace^* &= \delta^{\mu}_{\nu} \delta(\sigma_1 - \sigma_2) \,, \\
\lbrace \psi^a(\sigma_1), \psi^b(\sigma_2)\rbrace^* = i \eta^{ab} \delta(\sigma_1 - \sigma_2) \,, &\qquad
\lbrace \tilde\psi^a(\sigma_1), \tilde\psi^b(\sigma_2) \rbrace^* = i \eta^{ab} \delta(\sigma_1 - \sigma_2) \,.
\end{split}
\ee
All brackets between bosonic variables and fermionic variables with flat indices vanish. We now work out the algebra of constraints
\cite{Das:1988zk}, again using smeared constraints. We use Grassmann odd test functions for the Grassmann odd constraints
\be
\begin{split}
\{ \mathcal{Q}(\alpha), \mathcal{Q}(\beta) \}^* = i \mathcal{H}(\beta\alpha) \,, &\qquad \{ \tilde{\mathcal{Q}}(\alpha), \tilde{\mathcal{Q}}(\beta) \}^* = i \tilde{\mathcal{H}}(\beta\alpha) \,, \\
\{ \mathcal{Q}(\alpha), \mathcal{H}(\beta) \}^* = \mathcal{Q}(\alpha'\beta - \frac{1}{2}\alpha\beta') \,, &\qquad \{ \tilde{\mathcal{Q}}(\alpha), \tilde{\mathcal{H}}(\beta) \}^* = -\tilde{\mathcal{Q}}(\alpha'\beta - \frac{1}{2}\alpha\beta') \,, \\
\{ \mathcal{H}(\alpha), \mathcal{H}(\beta) \}^* = -\mathcal{H}(\alpha\beta'-\alpha'\beta) \,, &\qquad \{ \tilde{\mathcal{H}}(\alpha), \tilde{\mathcal{H}}(\beta) \}^* = \tilde{\mathcal{H}}(\alpha\beta'-\alpha'\beta) \,,
\end{split}
\ee
all brackets between tilded and untilded constraints vanish.

To package everything into representations of $O(D,D)$, we again make the substitution
\be
P \mapsto Y' \,,
\ee
and add the zero-mode term. This raises the same issues as in the bosonic case but there are no new complications: we do not double the fermions as they only transform under the action of $O(D,D)$ when contracted with a vielbein. We in fact introduce two double-vielbeine, $L_M{}^a$ and $R_M{}^a$ 
\cite{
Siegel:1993th,Siegel:1993xq, Hohm:2010xe,
Jeon:2011kp, Jeon:2011cn, Jeon:2011vx,  Jeon:2011sq,Jeon:2012kd,Jeon:2012hp} such that
\be
L_M{}^aL_N{}^b\eta_{ab} = \Pi_{MN} \equiv \frac{1}{2}(H_{MN}-\eta_{MN}) \,, \qquad R_M{}^aR_N{}^b\eta_{ab} = \tilde{\Pi}_{MN} \equiv \frac{1}{2}(H_{MN}+\eta_{MN}) \,.
\ee
These vielbeine have an $O(D,D)$ index $M, N, \ldots$ which like all $O(D,D)$ indices we lower and raise by the generalised metric, $H_{MN}$, as well as a flat $O(d-1,1)$ index $a, b, \ldots$ which is lowered/raised by the Minkowski metric $\eta_{ab}$.\footnote{The flat indices actually transform under different copies of $O(d-1,1)$
\cite{Hohm:2010xe,Jeon:2011kp, Jeon:2011cn, Jeon:2011vx,  Jeon:2011sq,Jeon:2012kd,Jeon:2012hp} 
and thus could be labelled by different indices. We do not use this convention to avoid the inevitable index clutter.} Furthermore, these vielbeine are eigenstates of the projectors onto left-/right-moving subspaces, $\Pi_{MN}, \tilde{\Pi}_{MN}$
\begin{equation}
 \begin{split}
 \Pi^{MN} L_N{}^a &= L^{Ma} \,, \qquad \Pi^{MN} R_{N}{}^a = 0\,, \\
 \tilde{\Pi}^{MN} R_N{}^a &= R^{Ma} \,, \qquad \tilde{\Pi}^{MN} L_N{}^a = 0\,,
 \end{split}
\end{equation}
and can thus be parameterised by spacetime fields as\footnote{Here $e_\mu{}^a$ and $\tilde
e_\mu{}^a$ are both vielbeine for the spacetime metric
\cite{
Siegel:1993th,Siegel:1993xq,Hohm:2010xe,
Jeon:2011kp, Jeon:2011cn, Jeon:2011vx,  Jeon:2011sq,Jeon:2012kd,Jeon:2012hp}.
 We define $\tilde \psi^\mu
= \tilde e^\mu{}_a \tilde\psi^a$, while $\psi^\mu = e^\mu{}_a\psi^a$.}
\begin{equation}
 L_M{}^a = \frac{1}{\sqrt{2}} \left( \begin{array}{c} e_{\mu a} - B_{\mu \nu} e^{\nu}{}_a \\ - e^\mu{}_a \end{array} \right) \,, \qquad
 R_M{}^a = \frac{1}{\sqrt{2}} \left( \begin{array}{c} \tilde e_{\mu a} + B_{\mu \nu} \tilde
e^{\nu}{}_a \\ \tilde e^\mu{}_a \end{array} \right). 
\end{equation}
We use these vielbeine to define
\be
\omega_{Mab} = L_{Na}\partial_ML^N\,_b + L^N\,_{[a}L^P\,_{b]}\partial_PH_{MN} \,, \qquad \tilde{\omega}_{Mab} = R_{Na}\partial_MR^N\,_b + R^N\,_{[a}R^P\,_{b]}\partial_PH_{MN} \,,
\ee
which are closely related to the normal spin-connections but are not themselves spin-connections as we will discuss a little further on.

With these objects we can rewrite the Hamiltonian action $O(D,D)$ covariantly
\bea
S &=& \int d\tau d\sigma \, \left( \frac{1}{2\pi}\dot{x}^M p_M + \frac{1}{2}\dot{\bar{X}}^M \eta_{MN} \bar{X}'^N - \frac{i}{2}(\psi^a\dot{\psi}^b + \tilde{\psi}^a\dot{\tilde{\psi}}^b)\eta_{ab} \right. \nonumber \\ && \left.  -\lambda\mathcal{H} - \tilde{\lambda}\tilde{\mathcal{H}} - i\xi\mathcal{Q} - i\tilde{\xi}\tilde{\mathcal{Q}} \right) \,,
\label{eq:doubleRNS}
\eea
where
\begin{equation}
\begin{split}
2\mathcal{H} &= X'^M \Pi_{MN} X'^N + i\eta_{ab}\psi^a\psi'^b + iX'^M(\tilde{\Pi}^{N}{}_M\omega_{Nab}\psi^a\psi^b - \Pi^{N}{}_M\tilde{\omega}_{Nab}\tilde{\psi}^a\tilde{\psi}^b)  \\
& -\frac{1}{4}\tilde{\Pi}^{MN}\omega_{Mab}\omega_{Ncd}\psi^a\psi^b\psi^c\psi^d -\frac{1}{4}\Pi^{MN}\tilde{\omega}_{Mab}\tilde{\omega}_{Ncd}\tilde{\psi}^a\tilde{\psi}^b\tilde{\psi}^c\tilde{\psi}^d  \\
& - \frac{1}{2}F_{abcd}\psi^a\psi^b\tilde{\psi}^c\tilde{\psi}^d \,, \\
2\tilde{\mathcal{H}} &= X'^M \tilde \Pi_{MN} X'^N - i\eta_{ab}\tilde{\psi}^a\tilde{\psi}'^b + iX'^M(\tilde{\Pi}^{N}{}_M\omega_{Nab}\psi^a\psi^b - \Pi^{N}{}_M\tilde{\omega}_{Nab}\tilde{\psi}^a\tilde{\psi}^b)\\
&-\frac{1}{4}\tilde{\Pi}^{MN}\omega_{Mab}\omega_{Ncd}\psi^a\psi^b\psi^c\psi^d -\frac{1}{4}\Pi^{MN}\tilde{\omega}_{Mab}\tilde{\omega}_{Ncd}\tilde{\psi}^a\tilde{\psi}^b\tilde{\psi}^c\tilde{\psi}^d  \\
& - \frac{1}{2}\tilde{F}_{abcd}\tilde{\psi}^a\tilde{\psi}^b\psi^c\psi^d \,, \\
\sqrt{2}\mathcal{Q} &= X'^M\eta_{MN}L^N{}_a\psi^a + \frac{i}{2}L^M{}_c \omega_{Mab}\psi^a\psi^b\psi^c + \frac{i}{2}L^M{}_c\tilde{\omega}_{Mab}\tilde{\psi}^a\tilde{\psi}^b\psi^c \,,\\
\sqrt{2}\tilde{\mathcal{Q}} &= X'^M\eta_{MN}R^N{}_a\tilde{\psi}^a + \frac{i}{2}R^M{}_c \omega_{Mab}\psi^a\psi^b\tilde{\psi}^c + \frac{i}{2}R^M{}_c\tilde{\omega}_{Mab}\tilde{\psi}^a\tilde{\psi}^b\tilde{\psi}^c \,,
\end{split}
\end{equation}
and we have defined
\begin{equation}
\begin{split}
F_{abcd} &= 2 L^M{}_{[a|} \partial_{M} (L^N{}_{|b]} \tilde{\omega}_{Ncd}) + 2L^M{}_{[a} L^N{}_{b]} \tilde{\omega}_{Mec}\tilde{\omega}_{Nd }{}^e 
 + 3 L^M{}_{[e|} \omega_{M|ab]} L^{Ne} \tilde\omega_{Ncd} \,,\\
\tilde{F}_{abcd} &= 2 R^M{}_{[a|} \partial_{M} (R^N{}_{|b]} \omega_{Ncd}) + 2R^M{}_{[a} R^N{}_{b]} \omega_{Mec}\omega_{Nd}{}^e{}
 + 3 R^M{}_{[e|} \tilde\omega_{M|ab]} R^{Ne} \omega_{Ncd} \,.\\
\end{split}
\end{equation}
For the undoubled string, the Hamiltonian constraints involved the Riemann curvature tensor and the
torsionful spin connections $\omega_{\pm \mu ab}$ \eqref{eq:tspinconnection2}. A natural question to
ask is whether the doubled action contains objects which have a geometric interpretation in double
field theory
\cite{Jeon:2011cn, Hohm:2011si,Geissbuhler:2011mx, Berman:2013uda,Geissbuhler:2013uka}. 

The objects $\omega_{Mab}$ and $\tilde \omega_{Mab}$ are connections allowing us to
covariantise the local $O(D) \times O(D)$ symmetry. They consist of a Weitzenb\"ock term
\cite{Geissbuhler:2011mx,Berman:2013uda, Geissbuhler:2013uka} plus a term
which is an $O(D)\times O(D)$ tensor. This extra term allows one to construct generalised
diffeomorphism scalars; $L^M{}_c \tilde \omega_{M ab}$
, $R^M{}_c \omega_{M ab}$, $L^M{}_{[c|} \omega_{M|ab]}$ and $R^M{}_{[c|}
\tilde\omega_{M|ab]}$ are generalised scalars, even though $\omega_{Mab}$ and $\tilde \omega_{Mab}$ themselves are not
generalised tensors. 

The above means that every term appearing in $Q, \tilde{Q}$,
$F_{abcd}$ and $\tilde F_{abcd}$ is a scalar under generalised diffeomorphisms. In addition, the
latter two consist of an $O(D)\times O(D)$ tensor plus a non-tensorial part:
\be
\begin{split}
F_{abcd} & = 2 L^M{}_{[a} \nabla_M ( L^N{}_{b]} \tilde \omega_{N cd} ) -2 L^M{}_{[a} L^N{}_{b]} \tilde
\omega_M{}^e{}_c \tilde \omega_{Nde} +  \Pi^{MN} \omega_{Mab} \tilde \omega_{Ncd} \,,\\
\tilde F_{abcd} & =2 R^M{}_{[a} \nabla_M ( R^N{}_{b]}  \omega_{N cd} ) - 2R^M{}_{[a} R^N{}_{b]} 
\omega_M{}^e{}_c \omega_{Nde} +  \tilde\Pi^{MN} \tilde \omega_{Mab}  \omega_{Ncd} \,,
\end{split}
\ee
where in the tensorial part the covariant derivative $\nabla_M$ acts on the $O(D)$ indices 
\be
\nabla_M L^N{}_a = \partial_M L^N{}_a - \omega_M{}^b{}_a L^N{}_b \,,\qquad
\nabla_M R^N{}_a = \partial_M R^N{}_a - \tilde\omega_M{}^b{}_a R^N{}_b \,.
\ee
One can also relate $\omega_{Mab}$ and $\tilde\omega_{Mab}$ to the ``semi-covariant'' derivative
\cite{Jeon:2011cn,Hohm:2011si} of double field theory, by appropriately projecting the latter. 

We now return to the analysis of the Hamiltonian system \eqref{eq:doubleRNS}. The Dirac brackets are
\be
\begin{split}
\lbrace X^M(\sigma_1), X^N(\sigma_2) \rbrace^* &= -\eta^{MN} \epsilon(\sigma_1 - \sigma_2) \,, \\
\lbrace \psi^a(\sigma_1), \psi^b(\sigma_2) \rbrace^* &= i \eta^{ab} \delta(\sigma_1 - \sigma_2) \,, \\
\lbrace \tilde\psi^a(\sigma_1), \tilde\psi^b(\sigma_2) \rbrace^* &= i \eta^{ab} \delta(\sigma_1 - \sigma_2) \,.
\end{split}
\ee
Now consider the algebra of constraints. The Dirac bracket of the $X^M$ implies that every bracket of constraints will generate many a priori non-zero terms of the form
\be
\lbrace F_1(\sigma_1), F_2(\sigma_2) \rbrace^* \,,
\ee
where $F_1$ and $F_2$ are combinations of background fields and their derivatives. These will prevent the algebra from closing, so we will demand
\be
\eta^{RS}\partial_R F_1(\sigma_1) \partial_S F_2(\sigma_2) = 0 \qquad \forall \sigma_1, \sigma_2 \,,
\ee
clearly analogous to the section condition in the bosonic case. This condition again requires the
background fields to be independent of half the coordinates and allows us to reduce to
\eqref{eq:rnssigmamodel} just as in the bosonic case. It is again possible that this condition could
be relaxed, an idea for further study \cite{Aldazabal:2011nj,Geissbuhler:2011mx,Geissbuhler:2013uka,Grana:2012rr,Berman:2013cli,Berman:2013uda}.

With this condition we can now study the algebra of constraints; a series of long and tedious calculations and judicious use of the Jacobi identity verifies that the algebra still holds:
\be
\begin{split}
\{ \mathcal{Q}(\alpha), \mathcal{Q}(\beta) \}^* = i \mathcal{H}(\beta\alpha) \,, &\qquad \{ \tilde{\mathcal{Q}}(\alpha), \tilde{\mathcal{Q}}(\beta) \}^* = i \tilde{\mathcal{H}}(\beta\alpha) \,, \\
\{ \mathcal{Q}(\alpha), \mathcal{H}(\beta) \}^* = \mathcal{Q}(\alpha'\beta - \frac{1}{2}\alpha\beta') \,, &\qquad \{ \tilde{\mathcal{Q}}(\alpha), \tilde{\mathcal{H}}(\beta) \}^* = -\tilde{\mathcal{Q}}(\alpha'\beta - \frac{1}{2}\alpha\beta') \,, \\
\{ \mathcal{H}(\alpha), \mathcal{H}(\beta) \}^* = -\mathcal{H}(\alpha\beta'-\alpha'\beta) \,, &\qquad \{ \tilde{\mathcal{H}}(\alpha), \tilde{\mathcal{H}}(\beta) \}^* = \tilde{\mathcal{H}}(\alpha\beta'-\alpha'\beta) \,,
\end{split}
\ee
with vanishing brackets between tilded and untilded constraints.

The first-class constraints $\mathcal{Q}, \mathcal{\tilde{Q}}, \mathcal{H}, \mathcal{\tilde{H}}$
generate symmetries of the action \eqref{eq:doubleRNS} up to the section condition. From their algebra we can identify $\mathcal{Q}$ and $\mathcal{\tilde{Q}}$ as supersymmetry generators under which the fields transform as
\begin{equation}
 \begin{split}
  \left\{ \mathcal{Q}(\alpha), X^M \right\}^* &= - \frac{1}{\sqrt{2}} \alpha L^M{}_a \psi^a + N^M(\mathcal{Q}(\alpha),X)\,, \\
  \left\{ \mathcal{Q}(\alpha), \psi^a \right\}^* &= \frac{1}{\sqrt{2}} \alpha \omega_{Mb}{}^a \psi^b L^M{}_c \psi^c - \frac{1}{2\sqrt{2}}\alpha L^{Ma} \omega_{Mbc} \psi^b \psi^c  \\ 
  & \quad - \frac{1}{2\sqrt{2}}\alpha L^{Ma} \tilde{\omega}_{Mbc} \tilde{\psi}^b \tilde{\psi}^c - \frac{i}{\sqrt{2}} \alpha X'^M L_{M}{}^{a}\,, \\
  \left\{ \mathcal{Q}(\alpha), \tilde{\psi}^{a} \right\}^* &= \frac{1}{\sqrt{2}} \alpha \tilde{\omega}_{Mb}{}^{a} \tilde{\psi}^{b} L^M{}_c \psi^c\,, 
\end{split}
\end{equation}
\begin{equation}
\begin{split}
  \left\{ \mathcal{\tilde{Q}}(\alpha), X^M \right\}^* &= - \frac{1}{\sqrt{2}} \alpha R^M{}_{a} \tilde{\psi}^{a} + N^M(\mathcal{\tilde{Q}}(\alpha),X)\,, \\
  \left\{ \mathcal{\tilde{Q}}(\alpha), \psi^{a} \right\}^* &= \frac{1}{\sqrt{2}} \alpha \omega_{Mb}{}^{a} \psi^b R^M{}_{{c}} \tilde{\psi}^{{c}}\,, \\
  \left\{ \mathcal{\tilde{Q}}(\alpha), \tilde{\psi}^{a} \right\}^* &= \frac{1}{\sqrt{2}} \alpha \tilde{\omega}_{Mb}{}^{a} \tilde{\psi}^{b} R^M{}_{c} \tilde{\psi}^{c} - \frac{1}{2\sqrt{2}}\alpha R^{Ma} \omega_{Mbc} \psi^b \psi^c  \\ 
  & \quad - \frac{1}{2\sqrt{2}}\alpha R^{Ma} \tilde{\omega}_{Mbc} \tilde{\psi}^b \tilde{\psi}^c + \frac{i}{\sqrt{2}} \alpha X'^M R_M{}^{a}\,,
 \end{split}
\end{equation}
where the non-local contributions are
\begin{equation}
 \begin{split}
  N^M(\mathcal{Q}(\alpha),X)(\sigma_1) &= - \frac{1}{\sqrt{2}} \int d\sigma_2 \eta^{MN}
\alpha(\sigma_2) \psi^a(\sigma_2) \left( \frac{i}{2} \partial_N \left( L^P{}_a \omega_{Pbc} \right) \psi^b \psi^c \right. \\
  & \quad \left. + \frac{i}{2} \partial_N \left( L^P{}_a \tilde{\omega}_{P{b}{c}} \right)
\tilde{\psi}^{{b}} \tilde{\psi}^{{c}} - X'^P \partial_N L_{Pa} \right)\! (\sigma_2) \epsilon(\sigma_1 - \sigma_2) \,, \\
  N^M(\mathcal{\tilde{Q}}(\alpha),X)(\sigma_1) &= - \frac{1}{\sqrt{2}} \int d\sigma_2 \eta^{MN}
\alpha(\sigma_2) \tilde{\psi}^{{a}}(\sigma_2) \left( \frac{i}{2} \partial_N \left( R^P{}_a \omega_{Pbc} \right) \psi^b \psi^c \right. \\
  & \left. \quad + \frac{i}{2} \partial_N \left( R^P_a \tilde{\omega}_{Pbc} \right) \tilde{\psi}^{b}
\tilde{\psi}^{c} + X'^P \partial_N R_{Pa} \right)\! (\sigma_2) \epsilon(\sigma_1 - \sigma_2) \,.
 \end{split}
\end{equation}
These supersymmetry transformations reduce to the usual ones upon undoubling \cite{Bergshoeff:1985qr}. Again, the section condition solves the problem of the non-local contributions. By requiring that the generalised metric only depends on $D$ coordinates and not on their $\eta_{MN}$-duals, we are guaranteed that the coordinates it does depend on always transform locally. The constraints $\mathcal{H}$ and $\mathcal{\tilde{H}}$ generate worldsheet reparametrisations of the fields. As for the bosonic case, the section condition is again needed to have a section where the coordinates transform locally.

One can pick superconformal gauge, $\lambda = \tilde{\lambda} = 1, \, \xi = \tilde{\xi} = 0$ to see
our action as a supersymmetrisation of the action in \cite{Tseytlin:1990nb}, up to the previously noted differences with bosonic zero-modes.

\section{Conclusions}

In this paper we have shown how the natural $O(D,D)$ invariance of the bosonic string Hamiltonian can
be used to construct an $O(D,D)$ invariant string action, which up to a subtlety with zero-modes is that found in the literature \cite{Tseytlin:1990nb}. In doing so we interpreted the $2D$ doubled coordinates of double field theory as the $2D$ coordinates of the string's phase space. There are some potential inequivalencies between our action and the normal string action. Firstly, there is an extra zero-mode term, which may have an effect quantum mechanically and should be studied in future work. Secondly, if the background fields are allowed to depend on all $2D$ coordinates, then the worldsheet diffeomorphism algebra fails to close. This provided a natural occurrence of the section condition as a way to restore worldsheet reparametrisation invariance, which we subsequently imposed.

This restricts the background fields to depend only on $D$ coordinates, but does so in an $O(D,D)$ covariant manner. In such a background we can identify $D$ coordinates as spacetime coordinates, and the other $D$ appear only through their $\sigma$-derivatives and can be interpreted as momenta; this reduces our action to the standard ones.

We then applied the same method to the RNS superstring, leading to the supersymmetrisation of the action found in \cite{Tseytlin:1990nb}. Only the bosonic degrees of freedom needed to be doubled as expected from a phase space point of view. We imposed the section condition on the background fields to ensure worldsheet parametrisation invariance.

The natural next step is to quantise this theory. We will have to investigate its equivalence with the
standard formulation of string theory; the total derivative term may play an important role
\cite{Hull:2006va, Berman:2007vi}. String backgrounds should appear out of massless modes of the
string, and there should be some way to generate only backgrounds which obey the section condition.
It will also be very interesting to see how T-duality switches between Type IIA and IIB string
theory \cite{Hohm:2011zr,Hohm:2011dv,Jeon:2012hp}. 

Another important avenue of future research is to investigate possible relaxations of the section
condition, perhaps to Scherk-Schwarz reductions
\cite{Aldazabal:2011nj,Geissbuhler:2011mx,Grana:2012rr,Berman:2013cli,Geissbuhler:2013uka,
Berman:2013uda}. The extension to non-trivial boundary conditions involving $O(D,D)$ twists is also
of interest as it would allow one to study the string in non-geometric backgrounds, where one expects to find modified Dirac brackets and
non-commutativity of physical coordinates \cite{Andriot:2012vb}.
Finally, the method presented in this paper might also lead to duality invariant actions for the Green-Schwarz string and perhaps even the membrane.

\section*{Acknowledgements}

We thank Malcolm Perry and Paul Townsend for many helpful discussions and comments on the
manuscript, and David Berman, Hadi Godazgar, Mahdi Godazgar and Jeong-Hyuck Park for other useful
discussions. CB is supported by the STFC, the Cambridge Home and EU Scholarship Scheme, St John's
College and the Robert Gardiner Memorial Scholarship. EM is supported by the STFC and a Peterhouse Research Studentship. AJR is supported by the STFC.

\bibliographystyle{JHEP}
\bibliography{NewBib}

\end{document}